# An Information System to support and monitor Clinical Trial Process

Daniela Luzi and Fabrizio Pecoraro

National Research Council. Institute for Research on Population and Social Policies, Rome, Italy
`d.luzi@irpps.cnr.it`

## ABSTRACT

*The demand of transparency of clinical research results, the need of accelerating the process of transferring innovation in the daily medical practice as well as assuring patient safety and product efficacy make it necessary to extend the functionality of traditional trial registries. These new systems should combine different functionalities to track the information exchange, support collaborative work, manage regulatory documents and monitor the entire clinical investigation (CIV) lifecycle. This is the approach used to develop MEDIS, a Medical Device Information System, described in this paper under the perspective of the business process, and the underlining architecture. Moreover, MEDIS was designed on the basis of Health Level 7 (HL7) v.3 standards and methodology to make it interoperable with similar registries, but also to facilitate information exchange between different health information systems.*

## KEYWORDS

*Clinical trial business process, Medical devices, System architecture, Data model, Trial registry.*

## 1. INTRODUCTION

Progress in clinical research depends largely on the results of Clinical Investigations (CIVs), which are complex processes encompassing different steps, from specification and planning to execution and final result analysis. There are an increasing number of applications that support CIV data management, project management and data quality control, thereby reducing time and costs and most importantly improving research quality. The demand for transparency in clinical research has further fostered the implementation of guiding principles that strongly encourage the registration of ongoing clinical studies in new developed national and/or international trial registries.

However, little attention has gone to information systems supporting both regulatory submission of CIV proposals and evaluation activities carried out by regulatory agencies and/or National Competent Authorities (NCAs) that have to approve CIV start and monitor their performance. These systems, that we consider different from trial registries, are designed to track the entire lifecycle of CIVs, from proposal submission to the collection of final investigational results. They enable CIV applicants to correctly observe national and international laws in the preparation and submission of a CIV proposal and help NCAs in the management of evaluation activities as well as in the monitoring of CIV performance. For these reasons we define these applications CIV monitoring systems. Their adoption can improve the communication flow between the different stakeholders, ensuring data quality, reducing time for CIV start, enhancing transparency of evaluation criteria and facilitating the monitoring of ongoing and concluded CIVs at national level. Such benefits also lead to decreased research costs, avoid duplication of CIVs and facilitate a timely placing of new clinical products into the market.

Moreover, a growing number of CIVs is required to test safety and efficacy of Medical Devices (MDs), which are becoming ever increasingly used tools in daily clinical practice to diagnose, treat and prevent many diseases. Given the differences between pharmaceutical products and MDs in terms of their development process, testing and commercialization [1], it is necessary to develop *ad hoc* systems that support the entire process of CIV on MDs, taking into account the characteristics of these wide range of products, specific CIV requirements and legal framework.

The paper describes a Medical Device Information System (MEDIS) developed by the National Research Council within a project supported by the Italian Ministry of Health. MEDIS combines the roles of both a registry of CIV data and a content repository of documents submitted by manufacturers to the NCA to obtain the approval for the clinical investigation start. In particular, MEDIS supports manufacturers in the documentation submission process as well as NCA evaluators in assessing the regulatory documentation received. It also manages the communications between the different stakeholders and collects data produced during the whole clinical investigation lifecycle, thus covering the functionalities of a CIV monitoring system. As described elsewhere [2], MEDIS design and development has been based on the application of Health Level 7 (HL7) [3] v.3 standards in order to develop a flexible and interoperable system.

After a concise overview of applications supporting clinical research (paragraph 2), the paper focuses on the main requirements of CIV monitoring systems comparing them to trial registries (paragraph 3). Based on these requirements, paragraph 4 provides a high level description of the CIV business process, while paragraph 5 reports the MEDIS data model. Paragraph 6 describes the MEDIS architecture, while paragraph 7 highlights examples of the main functionalities provided by MEDIS that help NCA tracking, managing and monitoring CIVs at national level.

## 2. APPLICATIONS SUPPORTING CLINICAL INVESTIGATIONS

Systems that support clinical investigations can be roughly subdivided into two categories. The first encompasses a variety of applications related to data and trial management systems (CDMS and CTMS) that support trial design [4] (e.g., protocol authoring systems, Case report form generation) as well as its performance (electronic data capture, patient recruitment, scheduling of clinical activities, site management, etc.) [5, 6]. In this context challenges remain regarding the adaptation of data models and applications in different clinical domains and pathologies as well as in various organizational frameworks. This often makes it necessary to develop tailored databases from scratch to manage new clinical studies. Moreover, the integration between these different legacy systems [7] is still a crucial issue, considering, on the one hand, the variety of both data and knowledge to be managed, and on the other the need to make these systems interoperable with external ones (Electronic medical records, laboratory systems, registries, etc.) [8-10]. In this context the use of HL7 messaging standards represents a suitable methodology to achieve system interoperability in the healthcare domain [11]. Other standard proposals aim to improve data exchange, such as the data model describing clinical protocols developed by the Clinical Data Interchange Standard Consortium (CDISC) that promotes platform-independent standardization to facilitate uniform data transactions [12,13].

The second category of systems developed in this domain concern trial registries, defined [14] as "a database of planned, ongoing or completed trials, published or unpublished, containing details of the trial's objectives, patient population, sample size and tested interventions". The demand for diffusing information on clinical studies already emerged in the mid '80s [15,16], underlining the need to overcome "publication bias" related to trials never published or published in sources difficult to access. In particular, attention was focused on the tendency not to publish trials with negative results, whose diffusion would enhance research transparency and also contribute to knowledge advancement [17,18]. Since then, different organizations have advocated mandatory submission of trial information [19] and a variety of trial registries have been developed [20-24]. At the same time important initiatives for implementation of a comprehensive registry have been undertaken, culminating in the development of the World Health Organization's (WHO) International Clinical Trial Registry Platform [25-27], that gathers information from important primary registries. The WHO initiative has had the merit of having established a 20-item minimum dataset as well as the adoption of the International Standardized Controlled Trial Number (ISRCTN) assigned by Current Controlled Trials [28].

At European level EUDRACT [29] (European Union Drug Regulating Authorities Clinical Trials) and EUDAMED [30] (European Databank on Medical Devices) provide information on clinical studies performed across European countries respectively on pharmaceutical products and MDs. The former was established in 2004 and recently provides a public version accessible via EudraPharm portal, while the latter has a restricted access to European NCAs and has become mandatory since May 2011. The advantages of these central registries are evident: uniform applications of the European Directives, unique information source, homogenous description of CIV data, and in case of MDs identification of a set of data that represent the wide variety of devices.

In fact, both databases have introduced a unique identifier that tracks the protocol irrespective of the clinical trial sites or member states involved. Both rely on the information provided by CIV applicants to each NCA, that then transmits the required data to the related European databases. To optimise this information flow, NCAs should develop their own local systems that support CIV information exchange with the European databases. These functionalities are fulfilled by MEDIS that foresees interoperability with other systems using a data model based on HL7 v.3. The adoption of local CIV monitoring systems will improve the entire CIV process, which generally still relies on paper-based acquisition of data and documents. To our knowledge, only the German Institute of Medical Documentation and Information has developed an information system that supports the process of CIV applications and facilitates the interaction between the German NCA and CIV applicants [31]. MEDIS system adds further functionalities to cover the entire CIV lifecycle also including serious adverse event reporting [32] and protocol amendments.

## 3. EXTENDING FUNCTIONALITIES OF TRIAL REGISTRIES

As previously introduced, systems developed by regulatory agencies and/or NCAs to collect and evaluate trial proposals as well as to monitor trial executions do not exactly fit in the category of trial registries. There is no common agreed definition for these systems; we call them CIV monitoring systems that in our view are hybrid applications that have to carry out a wider range of functionalities in line with NCAs' institutional role.

To summarise the main differences with trial registries a comparison is reported in table 1. Trial registries are generally freely available databases that diffuse information about clinical research to a vast and heterogeneous set of users, from investigators to the general public. They intend to improve research transparency [25], provide information for locating clinical trials [20], thus also preventing duplication of research studies and facilitating patient recruitment. On the other hand, CIV monitoring systems are additionally designed to manage data on CIV proposals to be evaluated before a CIV can start at the national study sites. They provide information generally restricted to an NCA's evaluation team and/or Ethical committees that access data and regulatory documentation to assess applicants' trial proposal. Another aim is to monitor trial executions establishing a communication flow between the stakeholders involved in whole CIV lifecycle. In the case of trial registries, the submission of trial information is strongly recommended for all interventional studies in human beings regardless of intervention type (consider the International Committee of medical Journal Editors' [28] policy, which requires registration as a precondition for publications). Conversely, in CIV monitoring systems the registration is a pre-requisite for the approval of the study. Another difference pertains to investigational products. Many trial registries consider both trial on pharmaceuticals and MDs [33], instead, CIV monitoring systems are generally managed by a specific authority in charge of and with specialised competences in assessing the deployment of the investigational product. For instance, MDs evaluation generally pertains to their safety requirements, while the investigation of pharmaceutical products has to ascertain their grade of toxicity [1]. From applicants' point of view, access to a CIV monitoring system is limited to their own proposals and/or ongoing studies, given the confidential nature of the information provided.

Table 1 – Comparison of trial registries and CIV monitoring systems.

| | Trial registries | CIV monitoring systems |
|---|---|---|
| Aim | <ul><li>Improving research transparency</li><li>Facilitating the location of clinical trials</li></ul> | <ul><li>Trial registries aims plus:</li><li>Management of CIV proposal to be evaluated</li><li>Trial monitoring</li></ul> |
| Target users | Investigators, industry, researchers, policymakers, general public | NCA's evaluation team, Ethical committees, CIV applicants |
| Access | Free | Restricted |
| Submission | Strongly encouraged | Mandatory |
| Investigational product | Drugs and/or devices | Devoted to either pharmaceutical products or devices |
| Information content | Database containing data mainly derived from clinical protocols (20 WHO items) | <ul><li>Database containing data derived from different regulatory documents;</li><li>Repository of administrative and technical documents with legal value</li></ul> |
| Process described | Milestones of the investigation performance | Whole CIV process from proposal submission to the collection final research results |

All this makes the content of the two types of system different. Trial registries mainly contain information taken from the clinical protocol (trial sponsor, title of the study, health conditions, inclusion and exclusion criteria, generally compliant with the WHO's 20-item recommendation [25]). In order to evaluate a study proposal, an NCA's evaluation team relies instead on a wider set of information, which is generally required according to national and international regulations and verified within the documents presented. In the case of CIVs on MDs, this is done not only on the basis of the clinical protocol, but also considering the risk analysis document, the MD technical description, the investigator brochure, etc. Therefore data gathered in application forms should represent pointers to particular parts of the CIV technical documents in order to facilitate their evaluation. It is also worth mentioning that data required for MDs under investigation is generally quite detailed, providing information such as risk classification, device accessories, sterilisation, etc. In trial registries that contain both information on pharmaceutical products and MDs this type of information is generally not reported. Moreover, an evaluation team has to examine all the administrative documents that guarantee the correct handling of the study from patients' point of view, for instance patient's informed consent, insurance, etc. Therefore, besides differences in the database content, CIV monitoring systems also act as repositories archiving documents with legal value providing also content management functionalities. More detailed information about the process of the entire lifecycle of a CIV has also to be provided in a CIV monitoring system in order to monitor the study and eventually take steps to safeguard patient safety, for instance in the case of serious adverse events.

This general picture shows that CIV monitoring systems have to support the interaction between the main stakeholders during the whole CIV process and in particular the management of:

- Regulatory documents submitted in the notification phase, including their updated versions and considering their legal value;
- Further data and/or documents exchanged during the whole CIV lifecycle to be registered also for legal purposes;
- Workflow that identifies activities in which specific data are collected during the CIV lifecycle.
- Collaboration activities performed during the evaluation process.

## 4. TITLE CIV BUSINESS PROCESS

The design of the MEDIS system was based on analysis of the domain of clinical investigation on MDs in close collaboration with the Italian NCA in charge for the evaluation of CIV proposals. Moreover, the description of the business process also took into account MD national regulations, European Directives [34], ISO technical norms [35], as well as guidelines for good medical practice [36]. The use case diagram in UML notation depicted in figure 1 shows a high level description of the MEDIS business process, describing the participating actors as well as the activities carried out in each sub-process.

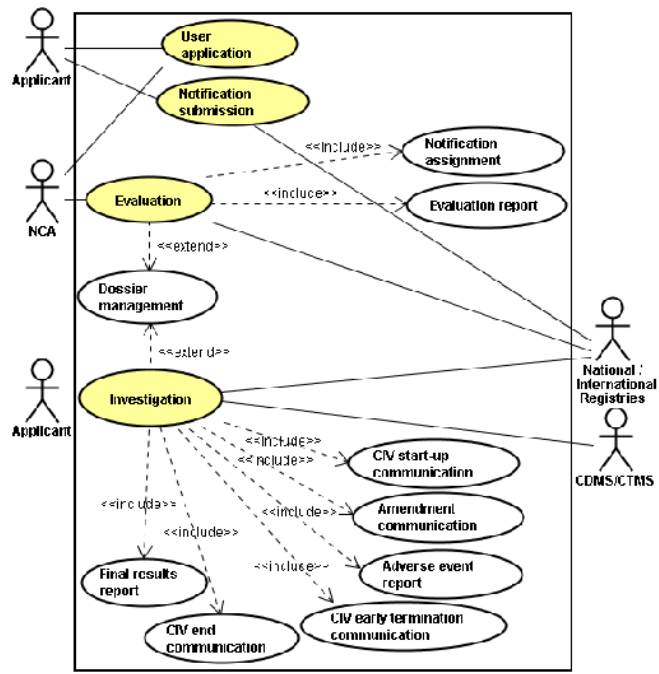

Figure 1.  Use case of MEDIS business process.

## 4.1. Actors

MEDIS system currently foresees two main user profiles:

- ▪ External users: applicants in the roles of manufacturer and/or authorized representative who submit a CIV proposal and communicate with NCA during the whole CIV lifecycle;
- ▪ Internal users: NCA personnel, who participate in different roles (supervisor, technical and medical evaluators) in the evaluation phase and monitor the whole CIV process.

Other participants could be included in the communication flow, such as Ethical Committees, that have to approve the CIV verifying its scientific soundness and ethical acceptability, and clinical investigators who could for example directly report information on serious adverse events. In addition MEDIS has been designed and developed to interact with other information systems, in particular with other national and international NCA registries, such as EUDAMED. This supports the information exchange of CIV on MDs among European NCAs, which is also one of the main demands of the new MD directive [34].

## 4.2. CIV lifecycle

The lifecycle of a CIV on MDs can be divided into four main sub-processes: 1) user application, 2) notification submission, 3) evaluation and 4) investigation.

- In the first sub-process (User application) applicants register to access the MEDIS system providing information about their organization (address, contact person, etc.) in addition to that related to the manufacturer(s) that might have delegated them to submit a study proposal. In the latter case the user has a profile of authorized representative in charge of CIV performance. The NCA assesses the data received and grants the user the right to access the system according to the identified profile(s).
- In the second sub-process (Notification submission), a CIV applicant fills in data and uploads regulatory documents to submit a new investigation proposal.
- The NCA receives the documentation and the evaluation sub-process can then start (Evaluation). This implies the assignment of the notification to an evaluation team (Notification assignment), generally a medical and a technical evaluator. In this sub-process the evaluation team assesses the MD safety requirements as well as the scientific, clinical and ethical fulfillments of the investigational plan. The evaluation sub-process ends with an Evaluation report, written by the NCA evaluation team, which approves or rejects the CIV start-up.
- If the CIV is approved, the Investigation sub-process can start. Its official inception, identified by the date of the first patient recruitment, is communicated by the applicant to the NCA (CIV start-up communication). During this sub-process the applicant is required to communicate further important information such as CIV end date and, if applicable, CIV early termination, amendments to the clinical protocol and/or serious adverse events. Finally, the applicant sends the Final report that gives the results of the CIV.

During the Evaluation and Investigation sub-processes, NCA can ask the applicant for further information and/or documentation. This information along with that collected during the whole CIV lifecycle are stored and managed in the dossier (Dossier management), which represents a central part of MEDIS information content.

## 5. CONCEPTUAL MODEL

Figure 2 shows MEDIS conceptual model using the UML class diagram notation at a high level of description. It outlines the main entities involved in the CIV process as well as the relationship between them. The class *Dossier* is the container of data and documents collected during the whole CIV lifecycle. It is managed by NCA *Evaluators* and composed by: 1) a *Notification* that is a composition of regulatory *Documents* submitted by the *Applicant* (in the *role* of manufacturer and/or authorized representative) during the notification submission sub-process; 2) a set of *Communications* exchanged between the *Applicant* and the NCA *Evaluators*. Each *Communication* may have one or more attached *Documents* and is related to the class *Communication type*. Similarly, a *Document* may be associated with other documents (for instance, an amended clinical protocol is associated with the previously sent clinical protocol as well as with a document containing the list of amendments) and is related to the class *Document Type*. The *Dossier* is also related to the class *Clinical Investigation* performed at least at one *Investigational site* where an *Investigational Device* is tested. An *Investigational Device* can be either a *Medical Device* or a kit, which is a composition of different *Medical Devices* and/or *Components*. Moreover, MDs can release a pharmaceutical product described by the class *Drug*. To evaluate its efficacy, an *Investigational Device* can be compared with a *Comparator Product* such as another *Medical Device* or a *Drug*. Moreover, a *Medical Device* may already have obtained a *CE mark,* required for sale in Europe for an intended use. A CE marked MD can also be investigated, if a CIV concerns an MD with a different intended use. Similar MDs can be associated through the relation *Similarity* that links the *Investigational Device* with the description of an already commercialized MD.

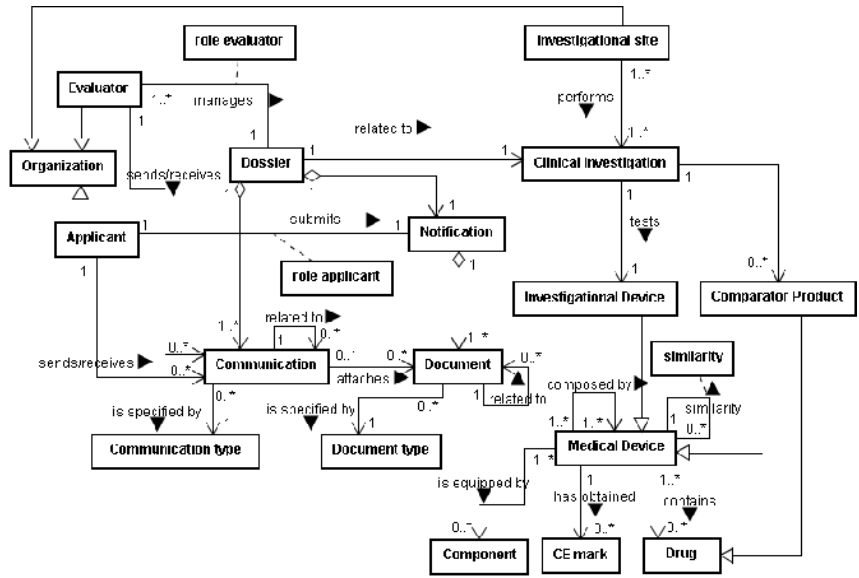

Figure 2. Conceptual model of the MEDIS system

Given the classes and relationships defined in the conceptual model we design the Domain Analysis Model (DAM) based on the HL7 v.3 standards. The use of HL7 and in particular the Reference Information Model (RIM) [3] foster already at an initial stage a common agreed data model that facilitates information sharing among heterogeneous systems and different organizations.

Previous works describe the MEDIS DAM [2], and the device domain model [37,38], while the serious adverse event communication model is reported in [32]. Moreover, the development of the MEDIS database has been based on a novel approach in which the DAM schema is converted into a logical data model based on the RIM [37,39].

The conceptual description of a MD in the context of CIVs represents a value added feature of the system design, as it details the different characteristics and roles of clinical products under investigation. This is evident if we compare this description with those released by HL7 that focus on specific MDs (i.e. cardiac implantable device) [40] or on data limited to track MDs in the context of surveillance [41]. Moreover, MEDIS MD domain model can be easily integrated in the CIV conceptual description provided by the BRIDG Project [13] that currently considers only pharmaceutical investigational products.

## 6. MEDIS ARCHITECTURE

From the architectural point of view, MEDIS is a client-server three-tiered system. The high level description of the MEDIS system architecture is depicted in figure 3 using the UML Component Diagram notation. All the components of the MEDIS system reside at the application logic layer and are developed adopting the Tapestry framework based on Java technologies such as JSP and Servlet. MEDIS presentation layer based on Web interfaces is composed by two web clients providing specific user interfaces respectively for applicants and NCA evaluators. This layer communicates with the application layer using HTTPS protocol that ensures secure transfer of reserved information.

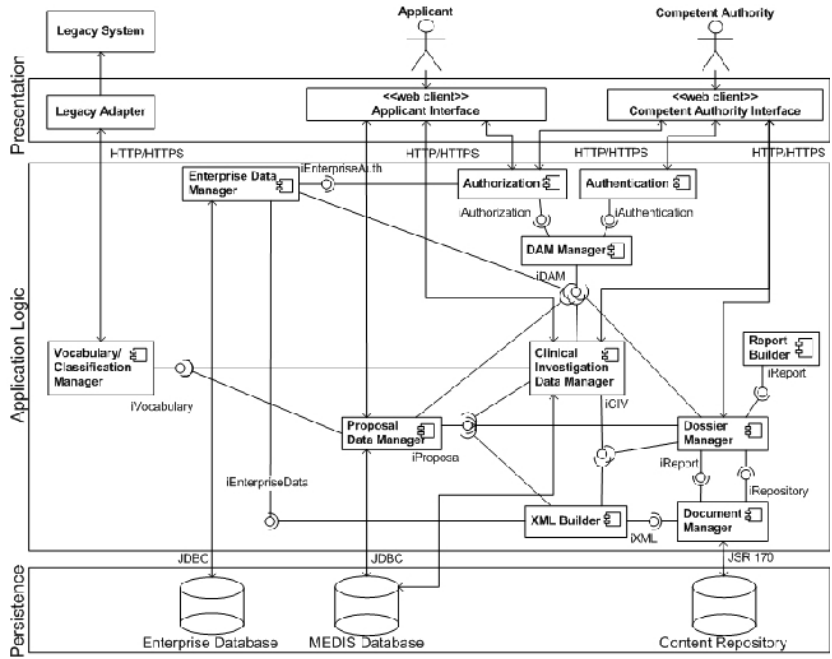

Figure 3. Architecture of the MEDIS system.

The application logic layer consists of a set of software modules handling the following functionalities:

1) The authentication of internal users is managed by MEDIS, while external users access the system via a single sign on portal based on the Central Authentication Service (CAS) project which controls their access to multiple and independent systems. Both internal and external users are authorized by MEDIS depending on their profiles. Each external user can access his/her portion of MEDIS system containing the list of submitted notifications together with those that are in the process of submission. Once a notification is submitted no change is allowed. The authorization of internal users depends on the role played in the evaluation of the CIV proposal (supervisor, medical/technical evaluator, administrative secretary). For instance each evaluator can enter and modify only the report he/she is in charge of, while the supervisor can access the evaluation report only if the evaluator allows him/her to do so. Based on the technical and medical reports, the supervisor can edit and send applicants the final evaluation of the CIV proposal.

2) The enterprise data manager supports information related to external users, the role they are allowed to play in the submission of notifications (manufacturer, authorized representative) as well as the data related to the organization responsible for the notification and/or delegated to do so. The administrative secretary accesses this system component and grants the users the right to initialize a new notification.

3) The DAM (Domain Analysis Model) component manages MEDIS conceptual model based on HL7 RIM (Reference Information Model). The design of MEDIS DAM and related data model are described in detail in previous works [2,32,37-39]. This component supports: a) the rules that validate the association between DAM objects; b) the rules that manage the workflow of the CIV lifecycle. It specifies the underlying workflow, determining the allowed steps depending on the state of the CIV lifecycle, thus enabling specific functionalities according to the CIV actual state as well as to the user profile. For instance this component manages the processes of notification of serious adverse events during the investigation sub-process.

4) The proposal data manager supports the acquisition of the notification, guiding applicants in the collection and submission of regulatory data and documents as well as verifying the completeness and consistency of data and documents submitted.

5) The CIV data manager supports the evaluation team in writing an evaluation report focused on crucial aspects such as MD characteristics, risk analysis and procedures planned for patient safety. This component also enables sharing the evaluation report among the evaluation team as well as editing the official communication of CIV approval or deny. Moreover, it supports the communication between applicants and NCA during the CIV performance, allowing applicants to notify important steps reached by an ongoing CIV such as starts, early termination or end data of the investigation. This component also manages the acquisition of the report that contains the CIV final results.

6) The Dossier Manager supports the storage and retrieval of documents uploaded as well as those created during the CIV lifecycle (i.e. NCA internal documents and communications). The XML component creates XML documents from the data provided in the electronic forms and then converts them into PDF (Report builder) in order to produce a digital signed document.

7) The vocabulary/classification manager connects MEDIS to external legacy systems to retrieve data such as vocabularies, classifications, and nomenclatures (i.e. MESH, MD repertoires, etc.).

Moreover, the above-described components use the following common functionalities:

a) The communication exchange that supports the information flow between applicants and the NCA evaluation team (e.g. request for data and/or document integration and their related replies), and also tracks the communication exchange.

b) The control of the completeness and consistency of the data and documents uploaded in a single form or in a correlated set of forms such as the initial and final report of serious adverse events.

c) The dynamic generation of the electronic forms depending on the user, the MD under investigation, and the step reached in the workflow, so that certain types of communication are allowed only in a specific phase of the CIV lifecycle.

d) The legal authentication of the documents and data submitted through a digital signature.

Finally, the MEDIS persistence layer is divided into:

1) A relational database containing information on applicants and their organization, as well as the organization delegating them to submit CIV proposals.

2) A relational database that contains data describing MD and CIV instances, data tracking the CIV lifecycle workflow and metadata related to documents stored in the content repository.

3) A content repository in charge of archiving documents uploaded as attachments of the notification or generated by the system starting with the data provided in the electronic forms.

## 7. SUPPORTING CIV MONITORING ACTIVITIES

To further facilitate CIV monitoring activities MEDIS provides functionalities that support an NCA in easily accessing data and documents exchanged within the entire lifecycle of the CIV. Monitoring activities related to an ongoing CIV are facilitated providing an overview of the exchanged documents, which outlines the time points at which they are produced and/or exchanged.

In figure 4 the upper table shows the list of documents exchanged during the CIV lifecycle, note that each document type is linked to the attached documents (Documenti allegati). In this case the notification identified by the code i.5.i.m.2/6/2009 (meaning the 6[th] notification of pre-market CIV received in the year 2009) gathers all regulatory documents submitted for the CIV approval. The lower table shows the list of requests and eventually the related responses given by the applicant. This interface allows users to select a specific document and also shows whether the communication activities have been fulfilled. In the case of figure 4 the user verifies whether a letter requiring further information on an early CIV termination received any reply and consequently takes further steps to obtain this information.

Figure 4. MEDIS interface accessing the documents contained in the dossier

Another functionality to help NCA monitoring activities is to facilitate the production of reports to evaluate the state of the art of CIV carried out at national level. MEDIS provides an advanced search (figure 5) that makes it possible to retrieve data on the number of ongoing and concluded CIVs (e.g. phase of the application process, types of applicants), on the types of MDs under investigation (e.g. risk class, characteristics and classification) and on CIVs (e.g. study design, population involved, multi-centric studies). In the specific case shown in figure 5, a search was carried out according to the following parameters: all the notifications submitted by a Manufacturer (Ruolo proponente = F) during 2009 (Anno = 2009) where CIVs were currently concluded (Stato notifica = sperimentazione conclusa).

Figure 5. MEDIS interface for browsing and searching information within the MEDIS system

## 8. CONCLUSIONS

MEDIS is a CIV monitoring system that extends the functionalities of trial registries. It enables both CIV applicants to correctly submit trial proposals and the NCA to evaluate them as well as monitor CIVs carried out at national level. Given MD increasing use in medical practice as well as their continuous incremental improvements in pre and post-market development, MEDIS represents an important tool for monitoring and exchanging information on CIVs. It assures data quality in terms of consistency and completeness of data submitted in the sub-process of CIV proposal, it improves the business process reducing the time for CIV start, additionally enhancing communication and monitoring activities.

MEDIS was designed on the basis of HL7 v.3 methodology and standards in order to make the system interoperable with other National registries and/or with EUDAMED. To further improve the functionality of the system, we intend to adopt HL7 CDA (Clinical Document Architecture) to standardise CIV documents specifying their structure and semantics. This would further improve document exchange as well as information retrieval of meaningful parts of these documents.
At the moment the MEDIS system has to be validated by real users in order to test its efficacy and efficiency as well as its performance. However, a successful introduction of MEDIS in the daily practice needs changes at an organisation level. NCA should gradually enforce the use of the system involving the different stakeholders that provide information of the CIV lifecycle and at the same time plan the necessary internal organizational changes to fully profit from the system that supports the new business process.

## ACKNOWLEDGEMENTS

This study was supported by the Italian Ministry of Health through the MEDIS project (MdS-CNR collaboration contract n° 1037/2007).

## Authors

Daniela Luzi is researcher of the Italian National Research Council at the Institute for research on populations and social policies. Her research interests have privileged the analysis of the impact of ICTs on information management and communication processes in healthcare. Research topics such as the analysis of clinical trial protocol under a semantic and structural perspectives, process modeling , HL7 standards have been dealt with in scientific papers and publications. She was the scientific responsible of the MEDIS project.

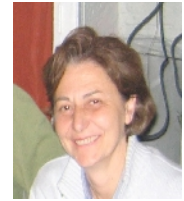

Fabrizio Pecoraro. Degree in Computer Engineering in Rome and Philosophy Doctorate in Bioengineering at the University of Bologna. During his doctorate studentship period he also held the position of assistant researcher at the University of Strathclyde, Glasgow, Scotland. Since 2007 he works as a researcher at the National Research Council – Institute of Research on Population and Social Studi es, Rome Italy – where his research activities mostly focus on the following aspects: business process analysis, development of conceptual models based on standard of clinical data such as HL7 and CDISC, design and development of information systems and definition of relational databases.

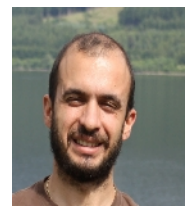